Growth and Anisotropic Properties of $RBa_2Cu_3O_x$ Single-Crystal Whiskers

Masanori Nagao[*], Satoshi Watauchi, Isao Tanaka, Takashi Okutsu[1], Yoshihiko Takano[1],

Takeshi Hatano[1], and Hiroshi Maeda[1]

University of Yamanashi, 7-32 Miyamae, Kofu, 400-8511, Japan

[1]National Institute for Materials Science, 1-2-1 Sengen, Tsukuba, 305-0047, Japan

Single-crystal whiskers of $RBa_2Cu_3O_x$ (R-123, R = Sm, Eu, Gd, Dy, Ho, Er, Tm, Yb, and Y) were successfully grown from Sb- or Te-doped precursors with a nominal composition of $R_{1.5}Ba_{2.75-3.00}Cu_3M_{0.5}O_x$ (M: Sb,Te). The whiskers were grown parallel to the *ab*-plane with the following typical dimensions: 2-8 mm in length, 20-80 μm in width, and 10-30 μm in thickness. In the case of rare-earth ions ($R^{3+}$) with a small ionic radius (Dy, Ho, Er, Tm, Yb, and Y), R-123 whiskers were grown from Te-doped precursors. The growths of Y-123 and Ho-123 whiskers were confirmed using both Te-doped and Sb-doped precursors. On the other hand, for rare-earth ions ($R^{3+}$) with a large ionic radius (Sm, Eu, and Gd), R-123 whiskers were grown only from Sb-doped precursors. Sb and Te were not detected in the grown whiskers by electron probe microanalysis. The anisotropies of R-123 single-crystal whiskers in the flux liquid state were estimated to be 6.0-33 from the angular dependence of their resistivities under various magnetic fields. The anisotropy increased with decreasing superconducting transition temperature.

*E-mail address: mnagao@yamanashi.ac.jp

## 1. Introduction

Since the discovery of high-$T_c$ cuprate superconductors such as Bi-based cuprates (Bi-2201, Bi-2212, and Bi-2223)[1] and $RBa_2Cu_3O_x$ (R-123, R = rare-earth elements)[2], much effort has been focused on the growth of the single crystals of these materials for both fundamental research and electronic applications. Single-crystal whiskers have become particularly attractive because of their useful characteristics such as perfect crystallinity, predictable growth directions, and excellent superconducting properties. Previously, we developed a whisker growth method using Te- and Ca-doped precursors, named the Te- & Ca-doping method, for Bi-based cuprates (Bi-2201, Bi-2212, and Bi-2223) and Ca-containing R-123[3-6]. Ca-doping into the precursors is required to enhance the growth rate of the whiskers by this method, but calcium ions contaminate the grown whiskers. We succeeded in growing Ca-free Y-123 single-crystal whiskers by the Sb-doping[7] and Te-doping[8] methods. We attempted to grow Ca-free R-123 single-crystal whiskers to systematically investigate the superconducting properties of the rare-earth element series.

In this paper, we report on the growth of Ca-free R-123 (R = Sm, Eu, Gd, Dy, Er, Tm, Yb, and Y) single-crystal whiskers by the Sb-doping and Te-doping methods and on the anisotropy ($\gamma_s$) of the grown whiskers. The relationship between rare-earth ion ($R^{3+}$) ionic radius and the growth conditions is examined. The superconducting transition temperature ($T_c$) dependence of $\gamma_s$ is also investigated.

## 2. Procedure

A precursor powder with a nominal composition of $R_{1.5}Ba_aCu_3M_{0.5}O_x$ ($a$=2.75 for M=Sb, $a$=3.00 for M=Te), with R = Sm, Eu, Gd, Dy, Ho, Er, Tm, Yb, and Y, was synthesized by a solid-state reaction as follows. We grew R-123 single-crystal whiskers with each R using Sb-doped and Te-doped precursors of compositions described in previous works[7-8]. The starting materials $R_2O_3$, $BaCO_3$, and CuO and the dopant $Sb_2O_3$ or $TeO_2$ were ground completely in a mortar, and calcined in air at 750-820 °C for 10 h. The grinding and calcination of the reactants were repeated three times. The calcined powder of 0.5 g was pressed into a pellet of 10 mm diameter and about 3 mm thickness. The precursor pellets were heated in air at $T_{max}$ °C for $t_{max}$ h, followed by slow cooling to 900 °C at a rate of $C_R$ °C/h, and then furnace-cooled to room temperature. $T_{max}$ and $t_{max}$ were varied between 920 and 1100 °C and 10 and 250 h, respectively, and optimized. After that, $C_R$ was optimized at the optimum $T_{max}$ and $t_{max}$. The crystal structure and composition of the as-grown whiskers were evaluated by X-ray diffraction (XRD) analysis using Cu Kα radiation and electron probe microanalysis (EPMA). The transport properties of the whiskers were measured by the standard four-probe method, and the $T_c$ was determined considering the criterion of a resistivity of 0.1 μΩ cm in resistivity-temperature ($\rho$-$T$) characteristics. $\Delta T_c$ was defined as the temperature difference between 10 and 90% resistivity. We measured the angular ($\theta$) dependence of resistivity ($\rho$) in the flux liquid state under various magnetic fields ($H$) and estimated $\gamma_s$ using the effective mass model[9-11].

## 3. Results and Discussion

Figure 1 shows a scanning electron microscopy (SEM) image of a typical as-grown Tm-123 whisker. The whisker exhibits a flat surface with a width of about 30 μm. R-123 whiskers except La-, Pr-, Nd-, and Lu-123 have a flat surface, as shown in Fig. 1. Figure 2 shows an XRD pattern of the flat surface of the Tm-123 whisker. The grown whiskers were confirmed to be of typical R-123 structure from the XRD patterns. The presence of sharp diffraction peaks of only the (00*l*) indices of the R-123 structure indicates that the flat surface is the *ab*-plane. The full width at half maximum (FWHM) of the grown whiskers was determined to be 0.07° from the (006) diffraction peak. We optimized the growth conditions to obtain longer whiskers of R-123 with various R (rare-earth) elements. Table I shows the optimum growth conditions such as the nominal compositions of the precursors, heat treatment parameters ($T_{max}$, $t_{max}$, and $C_R$), and lengths of the whiskers, where the $R^{3+}$ ionic radiuses are quoted from ref. 12. For $R^{3+}$ of small ionic radius (R = Dy, Ho, Er, Tm, Yb, and Y), R-123 whiskers were grown from the Te-doped precursors. Y-123 and Ho-123 whiskers were grown from both Te-doped precursors and Sb-doped precursors. On the other hand, for $R^{3+}$ of large ionic radius (R = Sm, Eu, and Gd), R-123 whiskers were grown only from Sb-doped precursors. The optimum heat treatment temperature ($T_{max}$) tended to increase for the whisker growth of R-123 as the $R^{3+}$ ionic radius increased. In the case of the Te-doping method, the whiskers tended to become short as the $R^{3+}$ ionic radius increased and were not obtained for $R^{3+}$ ionic radiuses larger than $Dy^{3+}$. The lengths of the grown whiskers were 2-8 mm. The obtained whiskers showed a flat *ab*-plane surface of 20-80 μm width and 10-30 μm thickness. The whiskers grown at a cooling rate ($C_R$) of 0.5 °C/h were longer than those grown at 1.0 °C/h, except for Yb-123, Dy-123, and Gd-123. The $T_{max}$ for $R^{3+}$ of larger ionic radius was higher than that for $R^{3+}$ of smaller ionic radius. Sb-doped precursors were effective for whisker growth at high temperatures, since a negligible amount of antimony oxide might have evaporated as compared with tellurium oxide during heating at high temperatures. As a result of qualitative analyses using EPMA, the concentrations of Sb and Te in the grown whiskers were found to be below the sensitivity limit of 0.1 wt %. Figure 3 shows the analyzed values of $u$ in $R_{1+u}Ba_{2-u}Cu_3O_x$ of the grown whiskers as a function of $R^{3+}$ ionic radius ($r_i$). For $R^{3+}$ with a smaller ionic radius than $Gd^{3+}$, $u$ shows a slight variation, and the compositions of grown whiskers are almost $RBa_2Cu_3O_x$. In contrast, for $R^{3+}$ with a larger ionic radius than $Dy^{3+}$, $u$ shows a significant variation from zero ($u=0$), and $u$ increases with increasing $r_i$. We suggest that the excess $R^{3+}$ ions substitute for the Ba sites ($Ba^{2+}$) in R-123 whiskers. The number of Ba site substitutions increases with increasing $r_i$, according to previous research on bulk materials[13]. This trend was also observed in Ca-containing R-123 single-crystal whiskers[6].

Figure 4 shows the $\rho$-$T$ characteristics parallel to the *ab*-plane for the Gd-123 and Er-123 as-grown whiskers. The $T_c$ values of the Gd-123 and Er-123 as-grown whiskers were 61 and 91 K, respectively. The $T_c$ of the Gd-123 as-grown whiskers was about 30 K lower than that of the Er-123 as-grown whiskers. The cause of the lower $T_c$ of the Gd-123 whiskers is the substitution of $Gd^{3+}$ ions for the $Ba^{2+}$ sites in Gd-123 whiskers and the oxygen deficiency in Gd-123 as-grown whiskers. It was reported that the $T_c$ of Ca-containing R-123 as-grown whiskers was markedly increased to about 80 K by annealing in oxygen[6,14]. Therefore, the lower $T_c$ of the Gd-123 as-grown whiskers is mainly due to the oxygen deficiency in Gd-123. The oxygen diffusion rate was suppressed, in which the $R^{3+}$ ions substitute for the Ba sites. One of the oxygen sites is totally vacant in the Cu-O chain layer and is located parallel to the Cu-O chains. Since the oxygen diffusion pass (CuO chain layer) is sandwiched by the Ba layers, the substituted $R^{3+}$ ions may trap oxygen and suppress the diffusion rate of oxygen in the whiskers during the furnace cooling to room temperature[6]. The $\Delta T_c$ values of Gd-123 and Er-123 were 3 and 2 K, respectively. The angular dependence of $\rho$ was measured at various $H$ in the flux liquid state to estimate the $\gamma_s$ of the grown whiskers, as introduced in refs. 9 and 10. Reduced field ($H_{red}$) is evaluated using the following equation for the effective mass model,

$$H_{red} = H(sin^2\theta + \gamma_s^{-2}cos^2\theta)^{1/2}, \qquad (1)$$

where $\theta$ is the angle between the *ab*-plane and the magnetic field[11]. $H_{red}$ is calculated from $H$ and $\theta$. $\gamma_s$ was estimated by the best scaling for $\rho$-$H_{red}$ relations. Figure 5 shows the angular dependence of $\rho$ at various magnetic fields ($H$ = 0.1-9.0 T) in the flux liquid state for Gd-123 and Er-123 single-crystal whiskers. The $\rho$-$\theta$ curves were represented by a two-fold symmetry. For Er-123 whiskers, small dips were observed in the $\rho$-$\theta$ curves around the $H$//*c*-axis at magnetic fields less than 1.0 T, as shown in Fig. 5 (b). A similar behavior was observed in Dy-123, Tm-123, and Y-123 whiskers. The small dip originated from twin boundaries in the whiskers, as was reported for a Y-123 single-crystal with twin boundaries[15]. On the other hand, for the whiskers of Sm-123, Eu-123, and Gd-123, the small dips were not observed in the $\rho$-$\theta$ curves. For $R^{3+}$ of larger ionic radius (R = Sm, Eu, and Gd), R-123 whiskers were free of twin-boundaries, since the whiskers were stabilized to a tetragonal structure by $R^{3+}$ substitution for the Ba sites in R-123 whiskers. Therefore, small dips were not observed for $R^{3+}$ of larger ionic radius (R = Sm, Eu, and Gd). The $\rho$-$H_{red}$ scaling obtained from the $\rho$-$\theta$ curves in Fig. 5 using eq. (1) is shown in Fig. 6. The $\gamma_s$ values of the Gd-123 and Er-123 whiskers were determined to be 13 and 8, respectively. Table II shows a summary of $T_c$ and $\gamma_s$ in R-123 whiskers. The $T_c$ in the as-grown whiskers of Dy-123, Ho-123, and Y-123 was dependent on the length of the whisker in each precursor pellet. $T_c$ in the short whiskers tends to be higher than that in the long whiskers. The $\gamma_s$ values of R-123 whiskers were 6.0-33. The decrease in $T_c$ is mainly due to the oxygen deficiency in Cu-O chain layers. Katayama *et al.* reported that the oxygen deficiency reduced the carrier concentration in R-123, so that $\gamma_s$ increases with decreasing carrier concentration[16]. Therefore, the $\gamma_s$ of Y-123 whiskers also increases with decreasing $T_c$. The plot of $\gamma_s$ vs $T_c$ obtained from Table II is shown in Fig. 7. For the whiskers with a close $T_c$, the $\gamma_s$ of R-123 whiskers with a larger $R^{3+}$ ionic radius (Sm, Eu, and Gd) was lower than that of R-123 whiskers with a smaller $R^{3+}$ ionic radius (Dy, Ho, Er, Tm, Yb, and Y). For example, the $\gamma_s$ values of the Y-123 with $T_c$: 43 K and Eu-123 with $T_c$: 42 K are 33 and 20, respectively. Two possible reasons for the lower $\gamma_s$ are considered. One is the substitution of the $R^{3+}$ ions for the Ba sites in the R-123 whiskers, and the other is the larger ionic radius of the $R^{3+}$ in R-123. Iwasaki *et al.* reported that the change in $\gamma_s$ for $x$ in $Nd_{1+x}Ba_{2-x}Cu_3O_y$ films was larger than compared with that in $Y_{1-x}Pr_xBa_2Cu_3O_y$ films[17]. Therefore, the lower $\gamma_s$ in R-123 with a large $R^{3+}$ ionic radius such as $Sm^{3+}$, $Eu^{3+}$, and $Gd^{3+}$ may be due to the substitution of $R^{3+}$ for Ba sites in R-123 whiskers. Further investigation is under way to clarify this phenomenon.

## 4. Conclusions

We have grown Ca-free R-123 single-crystal whiskers with R = Sm, Eu, Gd, Dy, Ho, Er, Tm, Yb, and Y using Sb- or Te-doped precursor pellets. R-123 single-crystal whiskers of 2-8 mm length were grown. In particular, in the case of R-123 single-crystal whiskers with larger $R^{3+}$ ions (Sm-123, Eu-123, and Gd-123), Sb-doped precursors were effective for R-123 whisker growth, and $R^{3+}$ ions were substituted partly in the Ba sites in R-123 whiskers. The increase in $\gamma_s$ tends to occur with decreasing $T_c$ of R-123 whiskers.

**Figure captions**

Fig. 1. SEM image of as-grown Tm-123 whisker.

Fig. 2. XRD pattern of the flat surface of Tm-123 whisker.

Fig. 3. Analyzed values of $u$ in $R_{1+u}Ba_{2-u}Cu_3O_x$ whiskers as a function of $R^{3+}$ ionic radius $r_i$.

Fig. 4. Resistivity-temperature ($\rho$-$T$) characteristics along the $ab$-plane of (a) Gd-123 and (b) Er-123 as-grown whiskers.

Fig. 5. Angular $\theta$ dependence of resistivity $\rho$ in flux liquid state at various magnetic fields ($H$ = 0.1-9.0 T) for (a) Gd-123 and (b) Er-123.

Fig. 6. Scaling of angular $\theta$ dependence of resistivity $\rho$ at a reduced magnetic field $H_{red}$ for (a) Gd-123 and (b) Er-123.

Fig. 7. Anisotropy $\gamma_s$ as a function of superconducting transition temperature $T_c$ for R-123 single-crystal whiskers.

Table I. Lengths of whiskers grown under optimum growth conditions, such as nominal compositions of precursors, heat treatment temperature ($T_{max}$), holding time at $T_{max}$ ($t_{max}$), and rate of cooling to 900 °C ($C_R$). $R^{3+}$ ionic radiuses are quoted from ref. 12.

| Nominal composition of $R_{1.5}Ba_aCu_{3.0}M_{0.5}O_x$ (M:Te,Sb) | | | Heat treatment conditions | | | Whiskers |
|---|---|---|---|---|---|---|
| R ($R^{3+}$ ionic radius: Å) | Ba:a | M | $T_{max}$ (°C) | $t_{max}$ (h) | $C_R$ (°C/h) | Length (mm) |
| Yb (0.858) | 3.00 | Te | 930 | 150 | 1.0 | 6-8 |
| Tm (0.869) | 3.00 | Te | 945 | 10 | 0.5 | 5-6 |
| Er (0.881) | 3.00 | Te | 965 | 10 | 0.5 | 5-6 |
| Y (0.892) | 2.75 | Sb | 995 | 30 | 0.5 | 3-5 |
| | 3.00 | Te | 1010 | 10 | 0.5 | 5-8 |
| Ho (0.894) | 2.75 | Sb | 1010 | 10 | 0.5 | 3-5 |
| | 3.00 | Te | 1030 | 10 | 1.0 | 2-4 |
| Dy (0.908) | 3.00 | Te | 1030 | 30 | 1.0 | 2-4 |
| Gd (0.938) | 2.75 | Sb | 1070 | 50 | 1.0 | 7-8 |
| | 3.00 | Te | 1000-1050 | 10 | 1.0 | ----- |
| Eu (0.950) | 2.75 | Sb | 1070 | 10 | 0.5 | 6-8 |
| | 3.00 | Te | 1010-1070 | 10 | 1.0 | ----- |
| Sm (0.964) | 2.75 | Sb | 1070 | 10 | 0.5 | 5-7 |

Table II. Superconducting transition temperatures ($T_c$) and anisotropies ($\gamma_s$) of R-123 single-crystal whiskers.

| R | Y | | | | Sm | Eu | Gd | Dy | | | Ho | | | Er | Tm | Yb |
|---|---|---|---|---|---|---|---|---|---|---|---|---|---|---|---|---|
| $T_c$ (K) | 43* | 55 | 58 | 87 | 39 | 42 | 61 | 71 | 81 | 60 | 78 | 82 | 91 | 92 | 84 |
| $\gamma_s$ | 33 | 28 | 18 | 7.5 | 17 | 20 | 13 | 12 | 9 | 20 | 11.5 | 10 | 8 | 7 | 6 |

*For Y-123 whiskers, lower $T_c$ samples ($T_c$: 43 K) were prepared by post annealing at 400 °C for 2 h in vacuum.

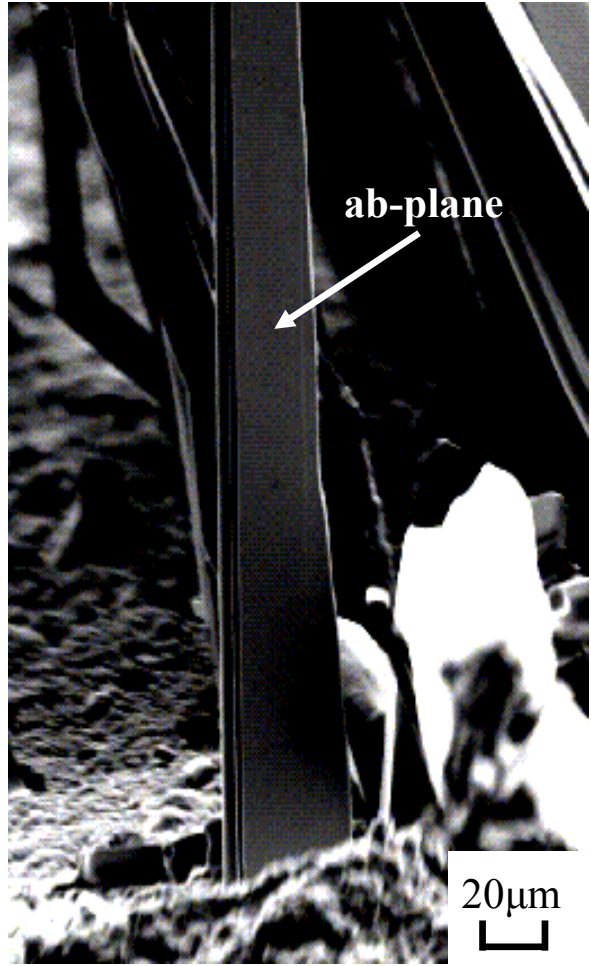

**Figure 1**

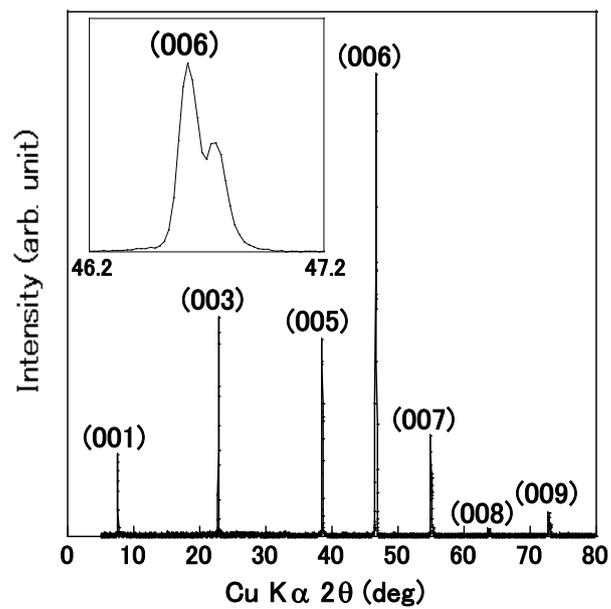

**Figure 2**

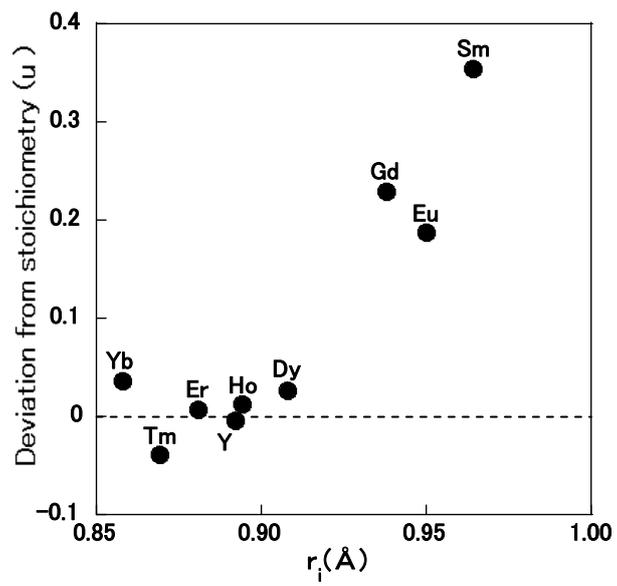

**Figure 3**

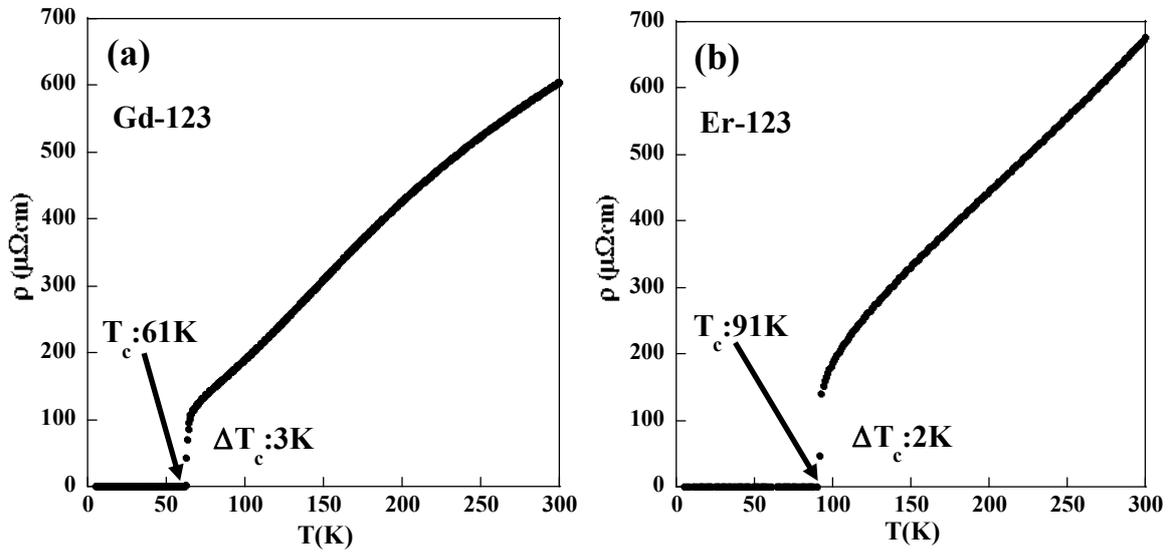

**Figure 4**

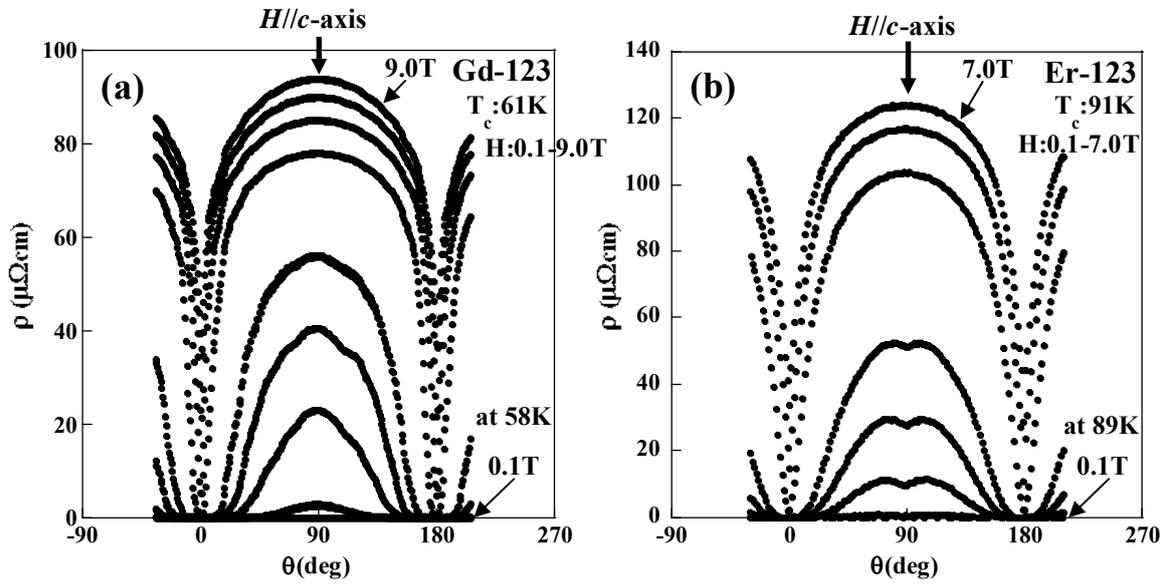

**Figure 5**

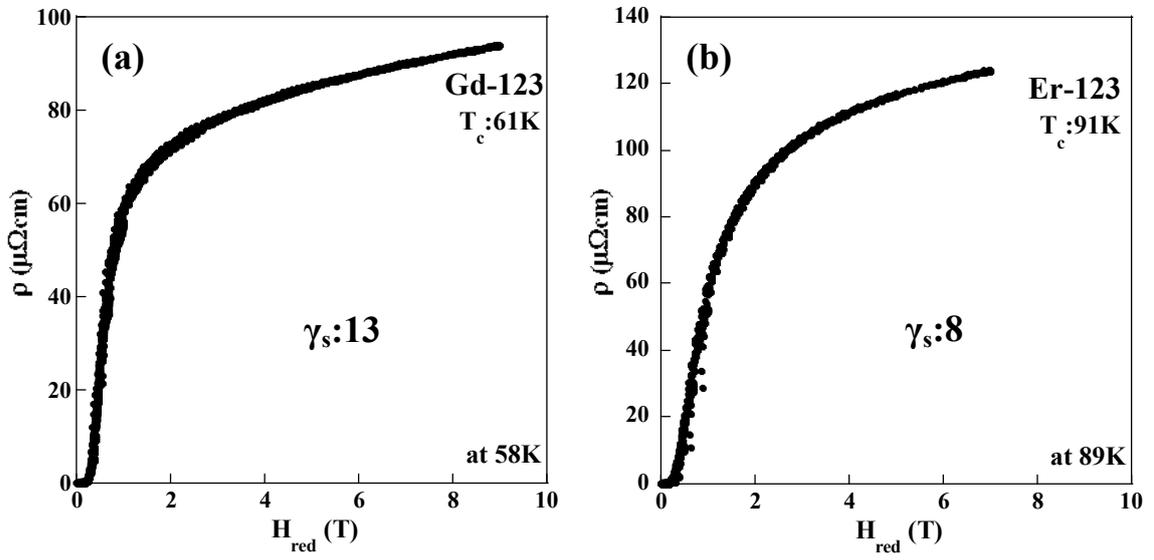

**Figure 6**

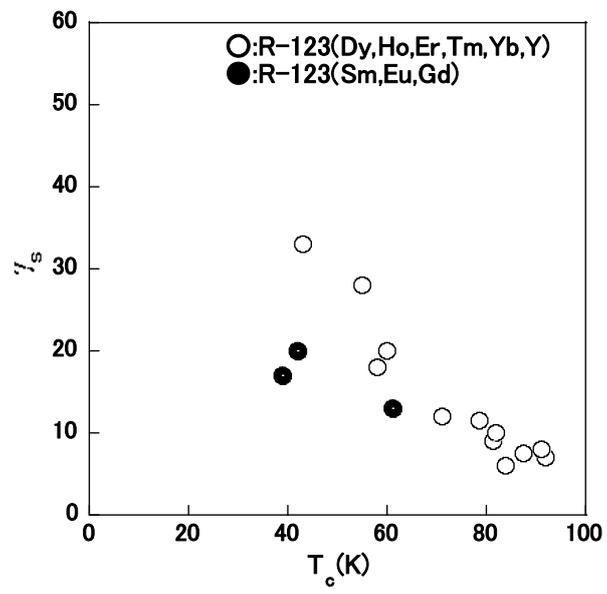

**Figure 7**